\documentstyle[aps,multicol,epsf,amstex]{revtex}
\begin{document}
\title{Superconductor-Quantum Dot-Superconductor
Junction in the Kondo Regime\\}
\author{Yshai Avishai
$^1$, Anatoly Golub$^1$ and Andrei D. Zaikin$^{2,3}$}
\address{$^1$ Ilse Katz Center for Nano-Technology \\
and \\
Department of Physics, Ben-Gurion University of the Negev,
Beer-Sheva, Israel\\
$^2$ Forschungszentrum Karlsruhe, Institut f\"ur Nanotechnologie,
76021 Karlsruhe, Germany\\
$^3$ I.E.Tamm Department of Theoretical Physics, P.N.Lebedev
Physics Institute, 117924 Moscow, Russia}
\maketitle
\date{Received 00,00 }
\begin{abstract}
Electron transport between two superconductors through an Anderson
impurity in the Kondo regime is investigated within the
slave boson mean field approximation. The current, shot noise
power and Fano factor are displayed versus the applied bias
voltage in the subgap region and found to be strongly dependent on
the ratio between the Kondo temperature $T_{K}$ and the superconducting
gap $\Delta$. In particular, the $I-V$ curve
exposes an excess current in
the limit $T_{K}/\Delta \gg 1$.
\end{abstract}

\begin{multicols}{2}

{\it Background}: A number of recently developed experimental techniques allow for
detailed investigations of electronic transport through atomic-size
metallic conductors\cite{Ru}. Usually, transport properties of such systems
are strongly affected by Coulomb
interactions. Novel physical effects emerge if electrodes of
an atomic-size contact become superconducting. In that case
the mechanism of multiple Andreev
reflections \cite{BTK} (MAR) plays a dominant role being responsible for both
dc Josephson effect and for dissipative currents at subgap voltages.
Further possibilities for experimental investigation of an interplay
between MAR and Coulomb effects in systems with few conducting
channels are provided by recently fabricated superconducting junctions
with a weak link formed by a carbon nanotube\cite{Kas,Christ}.

Recently we developed a theory\cite{us} for the study of
an $SAS$ junction consisting of an interacting quantum dot ($A$)
connected to superconducting ($S$) electrodes. It
enables the analysis of MAR in superconducting contacts with few
conducting channels in the presence
of electron-electron interactions.
An interplay between MAR and
Coulomb effects is responsible for novel effects
such as an interaction-induced shift of the subharmonic gap steps
on the $I-V$ curve and Coulomb blockade of MAR.
The latter may result in a strong suppression of the subgap
current through the dot at sufficiently low temperatures.

The combined effect of
MAR and electron-electron interactions on the shot noise\cite{rev}
in superconducting quantum dots was studied in Ref. \onlinecite{us1}.
It was demonstrated that interaction effects 
can strongly suppress the shot
noise power at subgap voltages. At the same time, the Fano factor
(proportional to the ratio of the shot noise power and the current)
was found to be nearly independent of interaction.

Our previous analysis\cite{us,us1} was restricted to physical
situations outside the Kondo regime implying that all relevant
energies in the problem were taken much higher than the Kondo
temperature $T_K \approx 0$. In the present work we study the
opposite limit of sufficiently {\it high} Kondo temperatures. This
case is relevant, e.g. for junctions composed of carbon nanotubes
\cite{Kas,Christ} where the Kondo effect with rather high Kondo
temperature $T_{K}\approx 1.6K$ was recently observed \cite{lin}.
Properties of superconducting quantum dots in this strong coupling
 regime are entirely distinct from
those exposed earlier in the limit $T_K \to 0$. On a qualitative
level, this difference between weak and strong coupling regimes can
be briefly summarized as follows: In the weak coupling limit (low
$T_K$), although the number of Andreev reflections $n\sim
2\Delta/eV$ may be large at low voltages $V$, both the current and
shot noise power remain rather weak. This is due to the low
effective transparency $\tilde\Gamma$ of the junction as a
consequence of strong repulsive electron-electron interaction
(Coulomb blockade). Large-$n$ processes are therefore damped as
$\tilde\Gamma^{n}$. By contrast, in the strong coupling Kondo
regime the effective transmission is much larger and a
ballistic-like channel opens up inside the dot. Hence, an
interplay between MAR  and the Kondo resonance is expected to
yield an excess current in the $I-V$ curve, similarly to the
case of noninteracting ballistic junctions. At very large values
of $T_K$ and in the low voltage limit this current should approach
the noninteracting result\cite{GZ} $I_{AR}=4e\Delta /h$.
Analogously, the shot noise power is expected to display a
pronounced maximum at $V=0$ and should decay as $1/V$ at small
bias as is familiar in the standard noninteracting $SNS$ junction
\cite{averin}. Below we will present a quantitative analysis which
fully supports this qualitative physical picture.

{\it Model and effective action}:
The pertinent system is represented by two half planar electrodes
($L$ and $R$) separated by the line $x=0$, and weakly
coupled to a point-like Anderson impurity $A$ located at the
origin. This model is of interest, for instance,
in connection with recent
experiments \cite{sara,kastner} on semiconductor quantum dots.
It was shown there that tunneling takes
place through a separate state with
features of a Kondo behavior (a tunable Kondo
effect).

The system dynamics is governed by the Hamiltonian
\begin{equation}
H=H_{L}+H_{R}+H_{d}+H_{t}+H_{c},
\label{H}
\end{equation}
in which $H_{L,R}$ are
the BCS Hamiltonians of the electrodes which depend on
the electron field operators $\psi_{L(R)\sigma}({\bf r},t)$
where ${\bf r}=(x,y)$ and $\sigma=\pm$ is the spin index.
As in Refs. \onlinecite{us,us1} the dot is described as a single level Anderson
impurity $A$ with energy $\epsilon_{0}<0$ and Hubbard repulsion
parameter $U$. In the Kondo regime of interest here we set $U \to
\infty$ and assume $|\epsilon_{0}|$ to exceed any other energy
scale except $U$. In this case it is convenient to express
the dot and the tunneling Hamiltonians $H_{d}$ and $H_{t}$ via
slave boson (operators $b,b^{\dagger}$) and
slave fermion (operators $c,c^{\dagger}$) auxiliary fields\cite{coleman}.
Explicitly, $H_{d}=\epsilon_{0} \sum_{\sigma}c_{\sigma}^{\dagger}
c_{\sigma}$ and $H_{t}={\cal T} \sum_{j\sigma}
c^{\dagger}_{\sigma} b\psi_{j \sigma}({\bf 0},t)+$h.c.,
where ${\cal T}$ is the tunneling amplitude. Finally, the
Hamiltonian of the system must also include a term which prevents
double occupancy in the limit $U\rightarrow\infty$. This
term reads
$H_{c}=\lambda(\sum_{\sigma}c^{\dagger}_{\sigma}c_{\sigma}+b^{\dagger}b-1)$,
where $\lambda$ is a Lagrange multiplier.

Let us now consider the dynamical ``partition
function''
\begin{equation}
Z \sim \int {\cal D}[F] \exp(i{\cal S}),
\end{equation}
where the
path integral is carried out over all fields $[F]$ and the
action ${\cal S}$ is obtained by integrating
the Lagrangian pertaining to the Hamiltonian (\ref{H}) along the Keldysh
contour. The procedure of
integrating out the electron fields
of the bulk electrodes $\psi_{L(R) \sigma}({\bf r},t)$ was described in details in \cite{us}. As a result we arrive at the
effective action expressed in terms of the Green functions of
the bulk superconductors. Our next step is to integrate out the
variables corresponding to the Fermi operators $c^{\dagger}_{\sigma}$
and $c_{\sigma}$ of the dot. The corresponding integral is
Gaussian, which yields,
\begin{eqnarray}
S_{\rm eff}=-i{\rm Tr} \ln\hat{G}^{-1}-
\int dt[\hat{\lambda}\sigma_{z}(\hat{b}\hat{b}-1)]. \label{Seff2}
\end{eqnarray}
Here $\hat{\lambda}=(\lambda_{1},\lambda_{2}) $,
$\hat{b}=(b_{1},b_{2})$ and $\sigma_{z}$ are diagonal matrices
acting in Keldysh space. Similarly to Ref. \onlinecite{us},
the inverse propagator $\hat{G}^{-1}$
depends on the Green functions of the electrodes.

{\it Mean field slave boson approximation (MFSBA)}: In order
to describe the Kondo regime we will treat the slave boson
fields  $b_{1}$ and $b_{2}$ in Eq.(\ref{Seff2}) within the
dynamical mean field approximation. Performing the
variation of the effective action with respect to $b_{1,2}$ and
$\lambda_{1,2}$ and then setting $b_{1}=b_{2}=b$ and
$\lambda_{1}$=$\lambda_{2}=\lambda$ we arrive at two
self-consistency equations that determine the
parameters $b$ and $\lambda$.

Before presenting these equations let us specify the expression
for the inverse propagator $\hat{G}^{-1}$. Performing
the standard basis rotation in Keldysh space one finds
\begin{eqnarray}
&&\hat{G}^{-1}(\epsilon,\epsilon')=\delta
(\epsilon-\epsilon')(\epsilon -\tau_z\tilde{\epsilon})+
\frac{\Gamma b^2}{2} \tau_{z}\hat g_{+}(\epsilon,\epsilon')\tau_z,
\label{Eq_Ginv}
\end{eqnarray}
where $\tilde{\epsilon} = \epsilon_{0}+\lambda$ is the renormalized
level position (in the Kondo limit one has $\tilde{\epsilon}\simeq 0$)
and $\Gamma \propto {\cal T}^{2}$ is the usual transparency parameter.
Here and below we define $\hat {g}_{\pm} = \hat {g}_{L}\pm\hat
{g}_{R}$, where
\begin{equation}
\hat g_{L,R}(t,t')=e^{\mp\frac{i\varphi(t)\tau_z}{2}}\int \hat g(\epsilon
)e^{-i\epsilon (t-t')}
\frac{d\epsilon}{2\pi}e^{\pm\frac{i\varphi(t')\tau_z}{2}},
\label{gL1}
\end{equation}
are Keldysh matrix Green functions of left and right
electrodes and $\dot \varphi /e=V(t)$ is the bias voltage across the dot.
The matrix $\hat{g}$ has the standard structure
with retarded and advanced Green functions
\begin{equation}
\hat g^{R/A}(\epsilon )= \frac{(\epsilon \pm i0)
+|\Delta|\tau_x}
{\sqrt{(\epsilon \pm i0)^2-|\Delta|^2}},
\label{Eq_g0RA}
\end{equation}
as diagonal elements $\hat{g}^{R/A}$ and the Keldysh function
$\hat g^K(\epsilon )= (\hat g^R(\epsilon )-\hat
g^A(\epsilon )) \tanh (\epsilon /2T)$ as the only nonzero (upper)
off-diagonal element. The Pauli matrices $\tau_{x,y,z}$
act in Nambu space.
The inverse of the matrix (\ref{Eq_Ginv}) is formally performed,
leading to a $2 \times 2$ Keldysh Green function with three elements,
\begin{eqnarray}
\hat{G}^{R,A}&=&[(i\frac{\partial}{\partial
t}-\tau_z\tilde{\epsilon})+
\frac{\Gamma b^2}{2} \tau_{z}\hat g_{+}^{R,A}\tau_z]^{-1}, \label {Eq GA}\\
\hat{G}^{K}&= & -\frac{\Gamma b^2}{2} \hat{G}^{R}\tau_{z}\hat
g_{+}^{K}\tau_z \hat{G}^{A}. \label {Eq GK}
\end{eqnarray}

In order to explicitly write down the self-consistency equations let us introduce the bare Kondo temperature $ T_{K}^{0}= D
exp[-\pi\|\epsilon_{0}|/(2\Gamma)]$ and define $ \Gamma
b^{2}=T_{K}^{0}X$, where $D$ is the energy bandwidth. Then our MFSBA
equations take the form
\begin{eqnarray}
X&=&-\frac{i\Gamma}{2T_K^0}{\rm Tr}\hat{G}^{K}\tau_{z}, \label{Eq x}\\
\lambda &=& \frac{i\Gamma}{8}{\rm Tr}[\hat{G}^{K}(g_{+}^{R}+g_{+}^{A})+
(\hat{G}^{R}+\hat{G}^{A})g_{+}^{K}] \label{Eq y}
\end{eqnarray}
where the trace also includes energy integration. Eq. (\ref{Eq x})
effectively determines the Kondo temperature (through the parameter $X$),
and reflects the constraint which prevents double
occupancy in the limit $U\rightarrow \infty$.
The second self-consistency equation (\ref{Eq y}) defines the renormalized
energy level position $\tilde{\epsilon}$.

Let us briefly discuss the validity range of the present
analysis. The MFSBA is known to encode the Kondo
Fermi-liquid behavior at low temperatures. An important parameter here
is the ratio between the Kondo temperature and the superconducting gap
$t_K \equiv T_{K}^{0}/\Delta$ \cite{fazio,raimondi,ambe}. For
$t_K\gtrsim 1$ a Fermi liquid behavior is expected.
Accordingly, in this regime Eq. (\ref{Eq x}) should have
a nonzero solution $X\neq 0$ which corresponds to nonzero $T_K$.
On the other hand, in the limit
of large $\Delta$ the only possible solution is a trivial one
$b=0$ (and, hence, $T_{K}=0$) \cite{hersh}. In this case --
as it was demonstrated in Ref. \cite{us} --
the problem can be treated within the dynamical mean field approximation for
the bare Anderson Hamiltonian. Quantitatively, the MFSBA is reliable
only for sufficiently large values of $t_K$. We believe, however, that
it can provide useful qualitative information also for moderate values
of $t_K$ describing a crossover between the Kondo regime and the
Coulomb blockade behavior \cite{us}.
It is worth noting here
that the applied bias voltage $V$ also attenuates
the Kondo resonance and lowers $T_{K}$. Hence, for the reliability of
the MFSBA in non-equilibrium situations, both $\Delta$
and $eV$ should not exceed the Kondo temperature.
Attention below is mainly focused on the subgap voltage regime $eV
\lesssim \Delta$ in which case $t_K$ appears to be the only relevant
parameter.

{\it I-V curve}: The standard
expression\cite{meir} for the tunneling current operator between the dot and
one (e.g. the right) electrode reads,
\begin{eqnarray}
I^{(1,2)}_{R} & = &\pm \frac{ie}{\hbar}\sum_{k}{ \cal T}
[\bar{c}\frac{1\pm \sigma_{x}}{2}\psi_{Rk}(0)-{\rm h.c.}]. \label{Eq_I12}
\end{eqnarray}
Here $\psi_{Rk}(0)$ is the Fourier transform of $\psi_{R}(0,y,t)$
with respect to $y$ and $(1,2)$ refer to Keldysh indices.
As before, it is convenient to integrate out
the $\psi$-fields and express the current $I$ through the dot in terms of the Green functions of the bulk electrodes. This procedure has been described in details in Ref. \cite{us}. As a result
we obtain
\begin{equation} \label{Eq_cur}
I=i\frac{eXt_K}{8\hbar}{\rm Tr}[(\hat{G}^{R}\tau_{z}-
\tau_{z}\hat{G}^{A})g^{K}_{-} -\hat{G}^{K}\tilde{g}], \label{I}
\end{equation}
where we defined $\tilde{g}
=\hat{g}^{R}_{-}\tau_{z}-\tau_{z}\hat{g}^{A}_{-}$ and
$\tilde{g}^{R,A}=\hat{g}_{-}^{R,A}\tau_{z}-\tau_{z}\hat{g}_{-}^{R,A}$.
Being combined with eqs. (\ref{Eq x}), (\ref{Eq y}) the result
(\ref{I}) can be conveniently used for computing the
transport current of an $SAS$ junction in the Kondo
regime for different values of $t_K$.

For sufficiently large $t_K$ we anticipate a strong Kondo resonance and the $I-V$ curve is expected to resemble that of purely
ballistic junctions without interaction\cite{GZ}. Indeed, in the
limit of large $t_K\gg 1$ which corresponds to the unitary, pure
ballistic case, our expression for the current (\ref{I}) reduces to 
that of Ref. \cite{GZ}. This agreement is further 
supported by our numerical analysis carried out for $t_K=100$
(dashed-dotted curve in Fig.1).  Calculation of
the current was also performed for $t_K=5$ and $1.6$ 
and the results are displayed in Fig. 1.
 \vskip  -1.0 truecm
\begin{center}
\leavevmode \epsfxsize=3.5in
\epsfbox{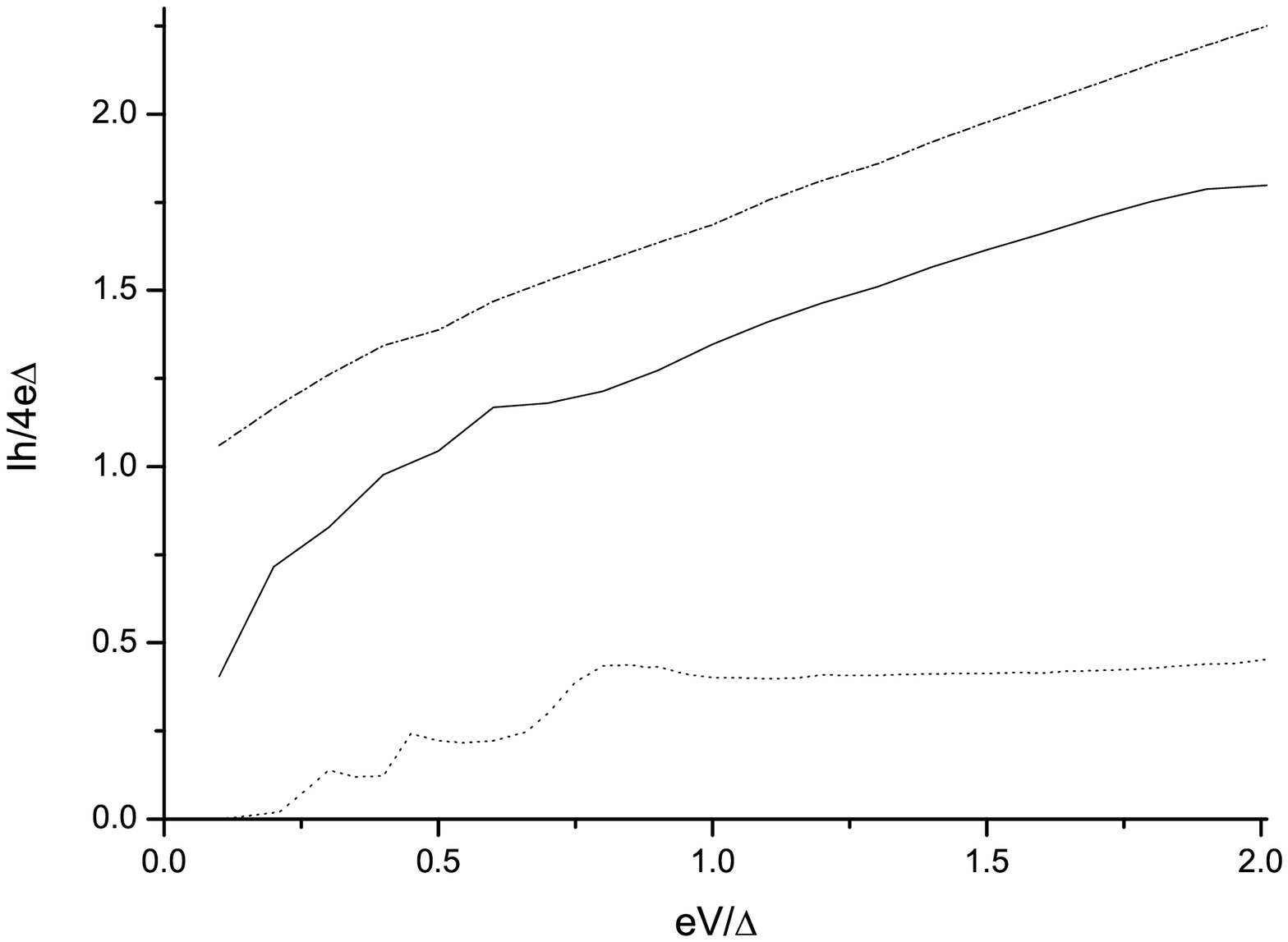}
\end{center}
\begin{small}
{\bf Fig. 1} The averaged current $I$ (in units
of $4e
\Delta/h$) versus the bias $V$ (in units of $\Delta /e$) for an $SAS$
junction at sub-gap voltages with
$\Gamma/T_{K}^{0}=200$. Dashed-dotted, solid and dotted curves correspond
respectively to  $t_K=100$, 5 and $1.6$.\\
\end{small}
\vskip -0.5 truecm 
For $t_K=5$, a pronounced excess
MAR current is clearly exposed in the $I-V$ curve, though its magnitude 
turns out to be smaller than in the unitary limit $t_K \gg 1$. 
One also observes sub-harmonic MAR steps (which are hardly
visible in the case $t_K \gg 1$). For a lower value 
$t_K=1.6$ the current is strongly suppressed (dotted line in Fig. 1), and the $I-V$ curve resembles that of a low transparency $SAS$. MAR steps in the $I-V$ 
curve become more pronounced as compared to the case of higher $t_K$. 

Let us briefly summarize our results for the current.
In the limit of large $t_K$ the $I-V$ curve is practically independent 
of $t_K$ and resembles that of a ballistic junction, as indicated in the upper curve in Fig.1. This effect is physically similar to that discussed in Refs.
\onlinecite{glazman,Tolya} where stimulation of the dc Josephson
current was found in the unitary limit of the Kondo regime.
For lower values of $t_K$ the junction is driven away from the 
unitary limit and the system behavior becomes much richer. It reflects the influence of both $\Delta$ and $V$ on the Kondo resonance and on the actual value of the Kondo temperature. While for $t_K=5$ some imprint of
ballistic behavior still remains (finite excess current), the
$I-V$ curve noticeably deviates from that obtained in a non-interacting limit. 
For $t_K=1.6$ the competition between gap-related suppression of the Kondo effectand the effective transparency of the junction becomes essential leading
to further decrease of the current. For even lower 
values $t_K$, the Kondo physics is no more relevant and one crosses 
over to the Coulomb blockade regime \cite{us}.

{\it Shot noise}: The shot noise spectrum is
usually defined as the symmetrized current-current correlation
function \cite{rev}. Being expressed via the
operators (\ref{Eq_I12}) it reads
\begin{eqnarray}&&
K(t_{1},t_{2})=\hbar [\langle \hat {T} I^{(1)}(t_{1})I^{(2)}(t_{2})\rangle
\nonumber \\
&&+\langle\hat {T} I^{(1)}(t_{2})I^{(2)}(t_{1})\rangle -
2\langle I\rangle^2], \label{Eq_K12}
\end{eqnarray}
where $\hat {T}$ is the time ordering operator and $\langle ...
\rangle$ denotes quantum averaging with the Hamiltonian
(\ref{H}). Substituting $I^{(1,2)}=(I^{(1,2)}_{L}-
I^{(1,2)}_{R})/2$ into eq. (\ref{Eq_K12}) we obtain an expression
for $K(t_{1},t_{2})$ which involves integration over surface
fields and dot electron slave particle field operators. The first
integration involving the Green function matrix $\hat {g}$ is
Gaussian and can be done exactly. Integrations over the dot slave
fermion fields is completed within the dynamic MFSBA. After
Fourier transform with respect to $t_{1}-t_{2}$ it is possible to
express the power noise spectrum $K(\omega)$ in terms of the Green functions of the entire system
(\ref{Eq GA}), (\ref{Eq GK}). It is convenient to decompose
$K=(K_{1}+K_{2})e^{2}\Delta/(8\hbar)$ with the result,
\begin{eqnarray}
K_{1}= \frac{Xt_K}{2} {\rm Tr} \{ (\hat{g}^{R}_{+}-\hat{g}^{A}_{+})
(\hat{G}^{R}-\hat{G}^{A})-\hat{g}_{+}^{K}\hat{G}^{K} \},
\label{Eq_K1}
\end{eqnarray}
\begin{eqnarray}&&
K_{2}=-\frac{(Xt_K)^2}{8}{\rm Tr} \{(\hat{G}^{K}\tilde{g})^{2}
-2\tau_{z}\tilde{g}\tau_{z} \hat{G}^{A}\tilde{g}\hat{G}^{R}-
\nonumber \\
&& [2\tilde{g}\hat{G}^{R}
\tau_{z}\hat{g}^{K}_{-}\hat{G}^{K}+(\tilde{g}^{R}\hat{G}^{R})^{2}-
(\hat{G}^{A}\hat{g}^{K}_{-}\tau_{z})^{2}+{\rm h.c.}] \}. \label{Eq_K2}
\end{eqnarray}

Expressions (\ref{Eq_K1}), (\ref{Eq_K2}) (supplemented by the
self-consistency eqs. (\ref{Eq x}), (\ref{Eq y})) are then solved
numerically for the same set of parameters 
$\Gamma/T_{K}^{0}=200$, $t_{K}=100$, 5 and $1.6$.
The results for the shot noise power spectrum $K$ versus the applied voltage 
$V$ are displayed in Fig. 2. These results are clearly correlated
with those for the $I-V$ curve and can be summarized as follows.

In the limit $t_K \gg 1$ our results are consistent with those
obtained for purely ballistic junctions \cite{averin}. In particular,
we mention that the noise spectrum $K$ exhibits a $1/V$ 
dependence at low voltages. At lower $t_K$ the physics is distinct. 
For $t_K=5$ (solid curve) the noise spectrum still shows features
typical for a junction with relatively high
transparency (see \cite{gu} and Fig. 1 therein), while
the results for $t_K=1.6$ (dotted curve) are more similar
to those for a low transparency junction.
For $t_K=5$ we get a pronounced noise power at low
voltage whereas at higher voltages $eV \gtrsim \Delta$ the shot
noise is significantly suppressed. At low voltage $K$ scales
approximately as $1/V$. The sequential tunneling picture is not valid 
in the unitary limit $t_K \gg 1$ as well as for the intermediate value
$t_K=5$ due to the interference between different $MAR$ processes.
Yet, this picture is partially restored at lower Kondo
temperatures, as is demonstrated by our results obtained for $t_K=1.6$.
\begin{center}
\leavevmode \epsfxsize=3.5in
\epsfbox{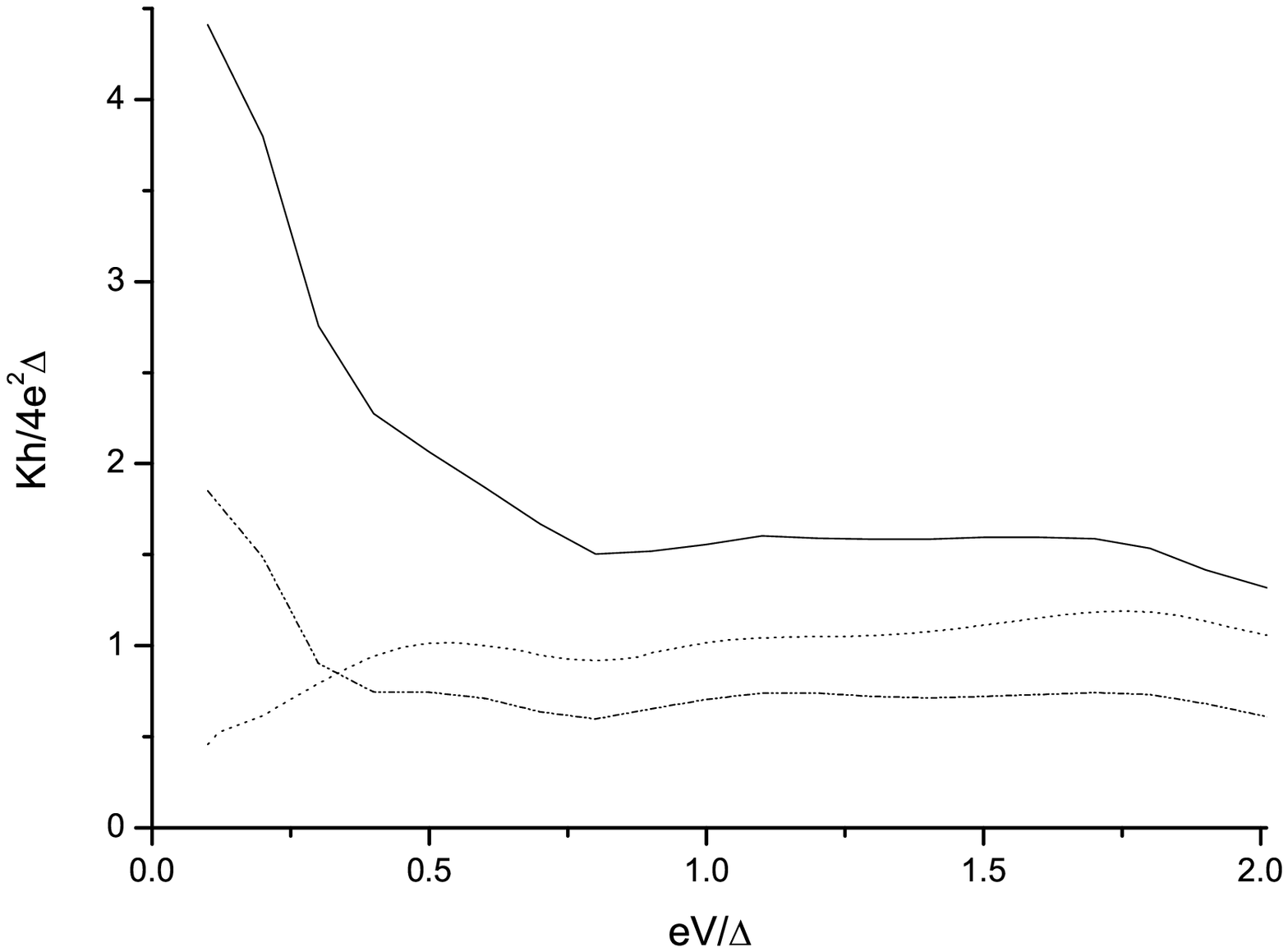}
\end{center}
\begin{small}
{\bf Fig. 2} The shot noise power $K$ (in units of $2e^{2} \Delta/h$)
as a function of $V$ (in units of $\Delta /e$) for an $SAS$ junction.
The parameters and notations are the same as in Fig. 1.\\
\end{small}
This effect of sequential tunneling due to $MAR$ is most clearly
manifested through the Fano factor $K/2eI$ depicted in Fig. 3 as a
function $V$.
\begin{center}
\leavevmode \epsfxsize=3.5in
\epsfbox{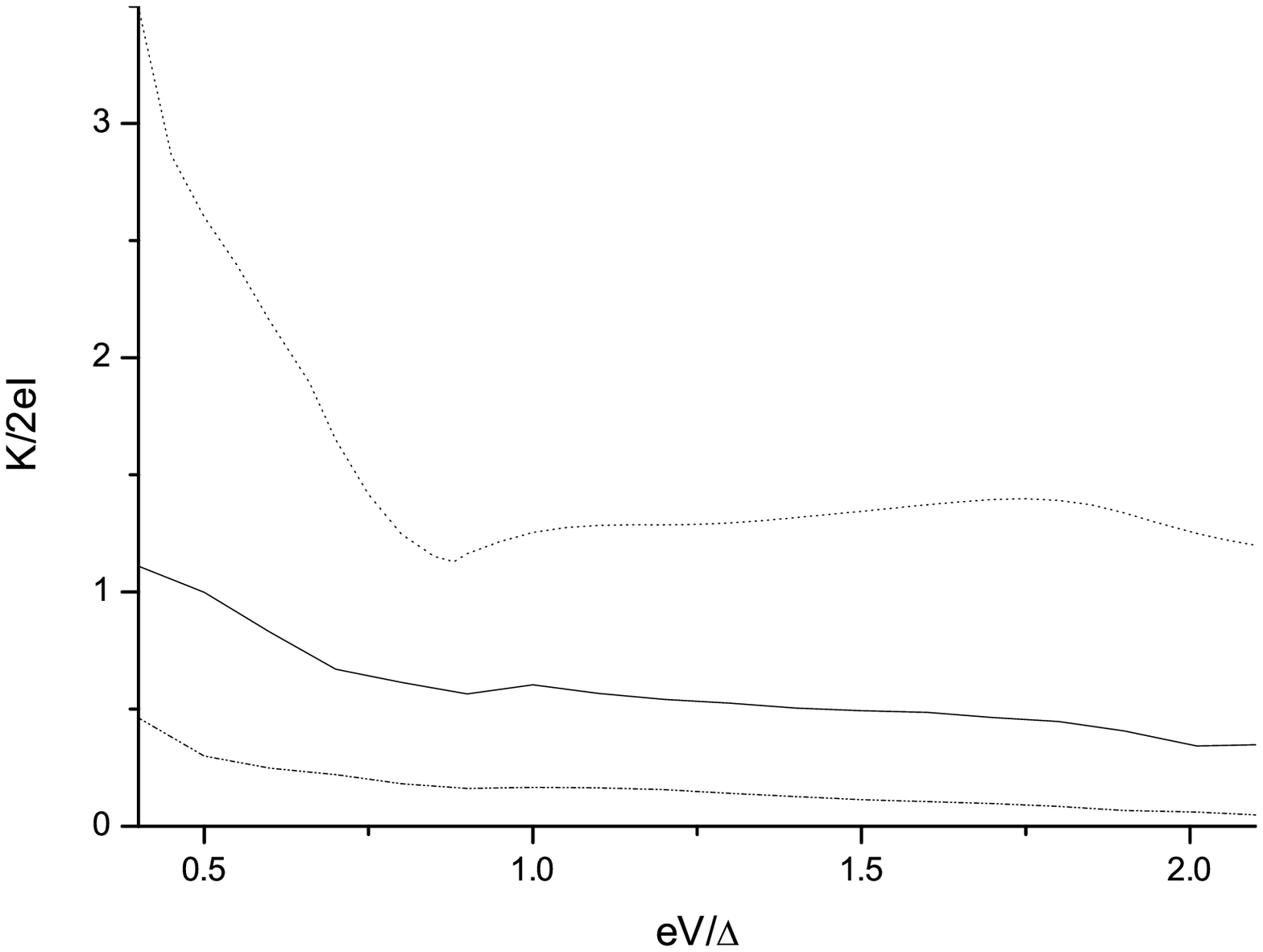}
\end{center}
\begin{small}
{\bf Fig. 3}
The Fano factor $K/2eI$ versus $V$. The parameters and
notations are the same as in Fig. 1. \\
\end{small}
\vskip -0.5 truecm
We observe that  for $t_K=5$ the Fano factor at
$eV\geq \Delta$ is substantially lower
than the expected value $K/2eI =2$  originated from Andreev
reflections. On the other hand, for $t_K=1.6$ the Fano factor is
closer to this value.

In conclusion, we have analyzed an important physical problem
involving strong correlations, the Kondo effect and
superconductivity. These aspects can be combined in an $SAS$
junction consisting of an Anderson impurity (in the Kondo regime)
located between two superconducting electrodes, which is
experimentally feasible. We have developed a
theoretical framework by which
it is possible to investigate an interplay
between MAR and Coulomb effects in the Kondo regime $T < \Delta < T_{K}$.
In this limit we have exposed the nonlinear $I-V$
characteristics and calculated the shot noise
power spectrum of $SAS$ junctions at subgap voltages $eV < 2 \Delta$.
We have found that at sufficiently large $t_K$ the Kondo resonance
plays the dominant role effectively making the junction behavior
similar to that of highly transparent
non-interacting weak links\cite{GZ,averin}.
This physical situation is qualitatively different from the Coulomb blockade
regime encountered in the limit $\Delta> T > T_{K}$
which we have analyzed in our previous works\cite{us,us1}.
A crossover between these two physically different regimes occurs
at $T_K \sim \Delta$ and is also -- at least qualitatively -- described
within our theoretical framework.

\noindent We would like to thank J.C. Cuevas, D.S. Golubev,
L.I. Glazman, J. Kroha and A. Rosch for discussions and useful suggestions.
This research is supported by DIP German Israel
Cooperation project {\bf Quantum electronics in low dimensions} by
the Israeli Science Foundation grant {\bf Many-Body effects
in non-linear tunneling} and by the US-Israel BSF grant {\bf
Dynamical instabilities in quantum dots}. This work is also a
part of the {\bf CFN} (Centre for Functional Nanostructures)
supported by the DFG (German Science Foundation).

\end{multicols}
\end{document}